\documentclass[]{emulateapj}

\shorttitle{The RINGS Survey I: H$\alpha$ and \ion{H}{1} Velocity Maps of Galaxy NGC 2280}
\shortauthors{Mitchell et al.}
\submitted{December 17, 2014}

\long\def\Ignore#1{{\relax}}
\def\eg{{\it e.g.,}}
\def\etal{{\it et al.}}

\newcommand{\favoredpa}{$157.7\pm0.4^{\circ}$}
\newcommand{\favoredeps}{$0.55\pm0.01$}
\newcommand{\favoredsys}{$1888\pm1\;$km~s$^{-1}$}
\newcommand{\favoredra}{6h44m49.09s$\pm$0.02s}
\newcommand{\favoreddec}{-27$^{\circ}$38\arcmin17.4\arcsec$\pm$0.4\arcsec}

\newcommand{\fppa}{$157.9\pm0.5^{\circ}$}
\newcommand{\fpeps}{$0.57\pm0.02$}
\newcommand{\fpsys}{$1887\pm2\;$km~s$^{-1}$}
\newcommand{\fpra}{6h44m49.09s$\pm$0.02s}
\newcommand{\fpdec}{-27$^{\circ}$38\arcmin17.7\arcsec$\pm$0.5\arcsec}

\newcommand{\radiopa}{$157.4\pm0.5^{\circ}$}
\newcommand{\radioeps}{$0.55\pm0.02$}
\newcommand{\radiosys}{$1888\pm1\;$km~s$^{-1}$}
\newcommand{\radiora}{6h44m49.03s$\pm$0.07s}
\newcommand{\radiodec}{-27$^{\circ}$38\arcmin16.2\arcsec$\pm$0.9\arcsec}

\begin{document}
\title{The RINGS Survey I:  H$\alpha$ and \ion{H}{1} Velocity Maps of
  Galaxy NGC 2280\altaffilmark{1}}

\author{Carl J. Mitchell, T. B. Williams\altaffilmark{2,3}}
\affil{Department of Physics and Astronomy, Rutgers University}
\affil{136 Frelinghuysen Road, Piscataway, NJ 08854}
\email{cmitchell@physics.rutgers.edu}
\email{williams@saao.ac.za}

\author{Kristine Spekkens, K. Lee-Waddell}
\affil{Department of Physics, Royal Military College of Canada}
\affil{P.O. Box 17000, Station Forces, Kingston, ON, K7K 7B4, XNS, Canada}
\email{kristine.spekkens@rmc.ca}
\email{karen.lee-waddell@rmc.ca}

\author{Rachel Kuzio de Naray}
\affil{Department of Physics and Astronomy, Georgia State University}
\affil{25 Park Place, Atlanta, GA 30303}
\email{kuzio@astro.gsu.edu}

\and

\author{J. A. Sellwood}
\affil{Department of Physics and Astronomy, Rutgers University}
\affil{136 Frelinghuysen Road, Piscataway, NJ 08854}
\email{sellwood@physics.rutgers.edu}

\altaffiltext{1}{Based in part on observations obtained with the
  Southern African Large Telescope (SALT) programme 2011-3-RU-003}

\altaffiltext{2}{South African Astronomical Observatory, Observatory,
  Cape Town 7925, South Africa}

\altaffiltext{3}{Astronomy Department, University of Cape Town,
  Rondebosch 7700, South Africa}

\begin{abstract}
Precise measurements of gas kinematics in the disk of a spiral galaxy
can be used to estimate its mass distribution.  The Southern African
Large Telescope (SALT) has a large collecting area and field of view,
and is equipped with a Fabry-P\'erot interferometer that can measure
gas kinematics in a galaxy from the H$\alpha$ line.  To take advantage
of this capability, we have constructed a sample of 19 nearby spiral
galaxies, the RSS Imaging and Spectroscopy Nearby Galaxy Survey
(RINGS), as targets for detailed study of their mass distributions and
have collected much of the needed data.  In this paper, we present
velocity maps produced from H$\alpha$ Fabry-P\'erot interferometry and
\ion{H}{1} aperture synthesis for one of these galaxies, NGC 2280, and
show that the two velocity measurements are generally in excellent
agreement.  Minor differences can mostly be attributed to the
different spatial distributions of the excited and neutral gas in this
galaxy, but we do detect some anomalous velocities in our H$\alpha$
velocity map of the kind that have previously been detected in other
galaxies.  Models produced from our two velocity maps agree well with
each other and our estimates of the systemic velocity and projection
angles confirm previous measurements of these quantities for NGC 2280.
\end{abstract}


\section{Introduction}
\label{intro}
The centrifugal balance of gas moving at close to the circular speed
in the plane of a disk galaxy offers a direct estimate of the central
gravitational attraction, and therefore a means to estimate the
distribution of mass within the galaxy \citep{Opik22, Robe69, TF77}.
The general flatness of spiral galaxy rotation curves at large radii
\citep{Babc39, Rubi80, Bosm81} provides some of the strongest evidence
for extended halos of dark matter (hereafter DM).

The standard Lambda Cold Dark Matter ($\Lambda$CDM) cosmological
paradigm is a highly successful model for the growth of structure in
the universe, particularly on large scales \citep{Spri06}.  However,
it is not yet clear whether galaxy formation within the $\Lambda$CDM
framework can produce systems whose underlying structure matches that
observed.  The clearest galaxy-scale prediction from the $\Lambda$CDM
model is the expected density profile of the DM halo, which was first
believed to have a simple broken power law form \citep{NFW96}.
However, a profile with a slope that decreases continuously inwards
from the break radius, where $\rho\propto r^{-2}$, and a central
density that is large but finite \citep{EH89}, appears to be a better
fit in DM only models \citep{Nava04, Nava10, Merr05, Gao08}.

Galaxies are believed to form in the DM halos as gas collects, cools,
and settles into rotational balance \citep{WR78, FE80, Gunn82, Kere05,
  DB06}.  The extra central attraction of the baryons in the inner
parts causes the DM halo to contract adiabatically, increasing its
inner density \citep{Blum86, Gned04, SM05}.  Furthermore, bursts of
star formation may release sufficient energy into the gas to cause it
to expand out of the proto-disk, and perhaps right out of the halo, on
a short time-scale.  Simulations of forming galaxies that include a
higher density threshold for star formation \citep[\eg][]{Gove10}
manifest repeated non-adiabatic changes in the disk mass that can
cause the halo density to decrease \citep[\eg][]{RG05, PG12} in models
of dwarf galaxies, but are less effective at higher masses
\citep{Broo11, Gued11, DiCi14, Pont14}.

Rotation curve measurements of spiral galaxies can, in principle, test
these gradually evolving predictions for the mass distribution.
However, a one-dimensional rotation curve does not contain sufficient
information to measure the separate distributions of dark and visible
matter \citep{vAS85, Sack97}.  No consensus has yet emerged on the
best way to separate the contributions of the stellar components from
that of the dark matter, \eg\ \citet{Wein01} and \citet{Bers11} reach
widely differing conclusions.  However, near maximal disks embedded in
lower density halos are favored by a number of theoretical arguments,
such as spiral-arm multiplicities \citep{SC84, ABP87} and bar-halo
friction \citep{DS00}.

As the baryonic mass in dwarf and low surface brightness galaxies is
believed to be a much smaller fraction of the total than in Milky Way
size galaxies, the rotation curve yields a more direct measure of the
DM halo density.  Various authors \citep[\eg][]{Rhee04, Haya06,
  Vale07} have raised a number of potential biases in such
measurements, such as non-circular motions, pressure and projection
effects, and halo triaxiality that need to be taken into account when
deriving the DM density profile.  These difficulties can be overcome
with good two-dimensional data, and \citet{KS11} find that the DM
density remains lower than predicted; baryonic processes have helped
to ease the tension \citep[\eg][]{Oh11, Gove12} but have not
eliminated it.  Other studies of the rotation curves of generally more
massive spiral galaxies \citep[\eg][]{McGa07} also suggest lower, and
more uniform, halo densities in the inner parts than predicted in
$\Lambda$CDM models.  In fact, \citet{Dutt07} suggest that the only
way to reconcile the predictions with the measured rotation curves of
larger galaxies is to invoke an unexplained de-compression of the DM
halo as the baryons settle to the center.

In order to address all these issues, we clearly require yet higher
quality measurements of velocities over the projected disks of
moderately inclined spiral galaxies.  Ideally, such measurements
should have the spatial resolution to determine the inner density and
extend to large enough radii to constrain the break radius of the halo
density profile.  Two-dimensional velocity maps are needed in order to
reveal non-axisymmetric flow patterns that could be missed in longslit
data, and modeling such flow patterns leads to improved estimates of
the mean interior density \citep{Spek07}.

The 11m Southern African Large Telescope (SALT) has a similar design
to the Hobby-Eberly telescope \citep{SR97}.  The Robert Stobie
spectrogaph (RSS) on SALT \citep{Buck06} is equipped with a
Fabry-P\'erot spectrophotometer (hereafter FP) that can measure the
Doppler shift of emission lines over the full 8 arcmin field of view.
A detailed overview of the SALT RSS Fabry-P\'erot instrument is
provided by \citet{Rang08}.

To take advantage of the large collecting area and field of view of
SALT, we have designed RINGS, the RSS Imaging and Spectroscopy Nearby
Galaxy Survey, that seeks to obtain detailed velocity maps at optical
spatial resolution over this large field of view for 19 nearby spiral
galaxies. \Ignore{We complement these data with lower spatial
  resolution, but more extensive, kinematic maps from aperture
  synthesis observations of the neutral hydrogen gas, together with
  multicolor photometry.}

A number of previous surveys have obtained two-dimensional velocity
maps of nearby galaxies with similar goals to those of our RINGS
program: \eg, \citet{Scho93}, BH$\alpha$BAR \citep{Hern05}, GHASP
\citep{Epin08}, GH$\alpha$FAS \citep{Hern08}, and DiskMass
\citep{Bers10}.  However, they had differing target selection
criteria, employed a variety of instrumental strategies and data
analysis methods, and were carried out on telescopes with apertures
ranging from 1.5-m to 4.2-m.  Some used optical spectroscopy only,
while others combined radio and optical data.

Our RINGS program uses a Fabry-Perot interferometer having an
8\arcmin\ field of view on a 10m optical telescope to produce very
deep and extensive kinematic maps with optical resolution.  We
complement these data with high-sensitivity, but lower spatial
resolution 21-cm aperture synthesis observations that provide
high-quality velocity measurements into the outskirts of the program
galaxies.  We also have new, deep multi-band photometry extending into
the near-infrared to constrain the luminous components of the
galaxies.  We will analyze these data with state-of-the-art dynamical
methods to produce the best possible determinations of the
distribution of luminous and dark matter within the target objects.
Our focus is on a relatively small sample of galaxies that nonetheless
covers a broad range of morphological structures and luminosities,
with an in-depth, handcrafted analysis of each galaxy that will allow
us to deal fully with the unique characteristics of each individual
target.  Our ultimate goal is to combine all these data in order to
model the mass distribution within all our program galaxies for
comparison with the predictions from $\Lambda$CDM cosmology.

\begin{deluxetable*}{cccc}
\tabletypesize{\footnotesize} \tablewidth{0pt} \tablecolumns{4}
\tablecaption{Galaxies Near NGC 2280 \label{table:neargals}}
\tablehead{\colhead{Galaxy Designation} &
           \colhead{Angular Separation} & 
           \colhead{Velocity Separation}&
           \colhead{Apparent R-band} \\
                            &
         \colhead{(arcmin)} &
         \colhead{(km~s$^{-1}$)}&
         \colhead{Magnitude\tablenotemark{a}}}
\startdata
ESO 427- G 005\tablenotemark{b} & 11.9 & $+136$ & 15.4\\
ESO 427- G 004\tablenotemark{b} & 13.3 & $-194$ & 15.8\\
AM 0643-272\tablenotemark{c}    & 14.4 & $+304$ & --  \\
ESO 490- G 035\tablenotemark{b} & 26.5 & $+242$ & 13.1\\
ESO 490- G 036\tablenotemark{b} & 29.3 & $+\phantom{0}39$ & 13.6\\
NGC 2272\tablenotemark{d}       & 30.3 & $+231$ & 11.3\\
AM 0644-280\tablenotemark{c}    & 31.6 & $+117$ & --  \\
ESO 490- G 044\tablenotemark{b} & 33.4 & $+269$ & 13.3\\
2MASX J06415402-2727239\tablenotemark{e} & 40.3 & $-\phantom{0}28$ & -- \\
ESO 490- G 031\tablenotemark{b} & 55.4 &  $+\phantom{0}92$ & 14.1\\
ESO 490- G 042\tablenotemark{b} & 56.8 & $-141$ & 14.6\\
ESO 490- G 038\tablenotemark{d} & 59.3 & $ +\phantom{0}75$ & 13.0\\
\enddata
\tablecomments{Galaxies with measured systemic velocies near that of
  NGC 2280 and angular separations from NGC 2280 less than
  $1^{\circ}$.  At our adopted distance to NGC 2280, an angular
  separation of 10\arcmin\ correspnds to a projected separation of
  68~kpc. The R-band magnitudes are to the $R_{25}$ isophote and that
  for NGC~2280 is 10.76.}
\tablenotetext{a}{\citep{LV89}} \tablenotetext{b}{\citep{Garc94}}
\tablenotetext{c}{\citep{Matt95}} \tablenotetext{d}{\citep{Huch12}}
\tablenotetext{e}{\citep{Patu05}}
\end{deluxetable*}

The purpose of this first paper is to demonstrate that the FP
instrument on SALT can provide reliable velocity maps of the quality
needed to accomplish the science goals of RINGS.  We therefore present
kinematic measurements of the H$\alpha$-line emission from excited
hydrogen in one of the first galaxies from the RINGS sample to be
observed with SALT, NGC 2280, and compare them with similar maps of
the same galaxy obtained using Karl G. Jansky Very Large Array (VLA)
observations of the 21cm line of \ion{H}{1}.  The spatial and velocity
resolutions of these two instruments differ substantially, and they
measure different components of the interstellar gas in the galaxy.
However, in both cases the radiating gas should be a good tracer of
the gravitational potential, and the measured velocities ought to
agree within the estimated uncertainties.  We find that this is indeed
the case.

\section{NGC 2280: Basic Properties}
\label{2280intro}
\begin{figure*}
\begin{center}
\includegraphics[width=.9\hsize]{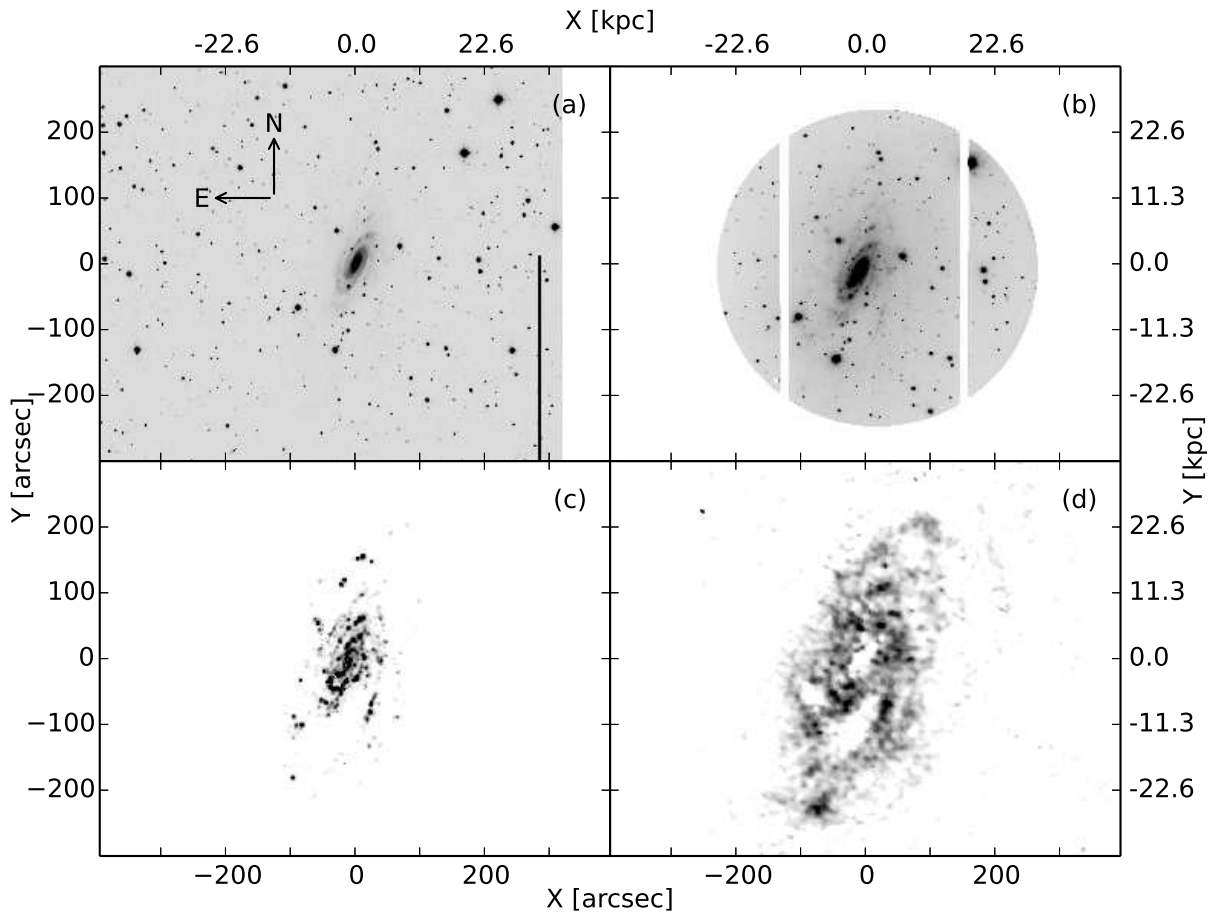}
\end{center}
\caption{Intensity maps of NGC 2280 with the same spatial scales. (a)
  An R-band continuum image acquired using the CTIO $0.9\;$m.  The
  indicated orientation is the same for all other panels.  (b) The
  stacked intensity of our SALT RSS Fabry-P\'erot images from 11
  Nov.\ 2011, which includes both the H$\alpha$ and continuuum
  emission integrated over the narrow wavelength range of our data.
  The circular boundary reflects the $\sim 8\arcmin$ field of view,
  and the gaps in the image are between CCD chips.  (c) The total
  H$\alpha$ intensity from line profile fits to our SALT Fabry-P\'erot
  data. (d) The total \ion{H}{1} intensity $\int I dv$ from the VLA.
\label{fig:inty}}
\end{figure*}

NGC 2280 is classified as a SA(s)cd galaxy \citep{deVa91}.
Figure~\ref{fig:inty}(a) shows an R-band continuum image taken with
the CTIO 0.9m telescope (Kuzio de Naray \etal, in
peparation). \citet{LV89} give a value for $R_{25}$ of 380\arcsec\ in
the B band.  \citet{Kori04} give a heliocentric systemic velocity of
$v_{\rm sys} \simeq 1899\;$km~s$^{-1}$.  The galactocentric velocity
of $v_{\rm gsr} \simeq 1703\;$km~s$^{-1}$, together with an adopted
Hubble constant of $H_0=73.0\;$km~s$^{-1}$~Mpc$^{-1}$, indicate a
distance of $\sim 23.3\;$Mpc, and thus an angular scale of
$0.113\;$kpc~arcsec$^{-1}$.  At this distance, the apparent V-band
magnitude, $m_V=9.61$ \citep{deVa91}, corresponds to an absolute
magnitude $M_V=-22.23$.  Its radio flux at $1.49\;$GHz is $59.7\;$mJy
\citep{Cond96}.

\citet{AM87} note the existence of five apparent companion galaxies to
NGC 2280.  We searched NED\footnote{\tt http://ned.ipac.caltech.edu}
for galaxies within one degree of the center of NGC~2280 having
redshifts within 500~km~s$^{-1}$ of its systemic velocity.  We list
these possible companion galaxies in Table~\ref{table:neargals},
however we do not find significant indications of recent interactions
or tidal distortions in our data.

\section{H$\alpha$ Fabry-P\'erot Observations}
\label{fp}
\subsection{Observations}
We have measured the H$\alpha$ emission from the disk of NGC 2280 using
the medium-resolution mode of the RSS Fabry-P\'erot etalon on SALT.
On each of two nights, 1 Nov.\ 2011 and 28 Dec.\ 2011, we obtained 25
one-minute exposures, covering the spectral region 6565.5\AA\ to
6643.5\AA\ in equally-spaced 2\AA\ steps.  In order to deal with
possible drifts of the etalon wavelength calibration, we first scanned
the spectral range in 4\AA\ steps, and then filled in with exposures
at interleaving spacings of the etalon.

The seeing for the nights of 1 Nov.\ 2011 and 28 Dec.\ was
approximately 1.75\arcsec\ and 2.0\arcsec, respectively.  The pixel
scale of the CCD is 0.5\arcsec.

Our observations were taken in single-etalon, medium spectral
resolution mode.  The RSS Fabry-P\'erot system is designed to operate
in dual-etalon mode, using a lower resolution etalon to reject
unwanted adjacent interference orders of the medium resolution etalon
\citep[see][for details]{Rang08}.  Unfortunately, the current
mechanical support structure of the etalons in the spectrograph does
not maintain their position with sufficient rigidity; misalignments
introduce reflections between the two etalons that give rise to a
series of strong ghost images, making dual-etalon observations
impossible.  Our single-etalon mode observations use only the medium
resolution etalon and an interference filter.  This allows not only
the desired interference order to be recorded, but also two other
adjacent orders transmitted by the filter.  This is not a serious
impediment for our measurements of emission lines, since there are no
other emission features at the other transmitted wavelengths
(separated by the 75~\AA\ free spectral range of the medium resolution
etalon).  The galaxy continuum and the night sky continuum are however
three times brighter than would be obtained from a single order,
reducing the signal-to-noise achieved in our measurements.  Planned
mechanical modifications to the spectrograph will allow dual-etalon
measurements to be made in the future.

\subsection{Photometric calibration}
We also obtained twilight sky flat-field images, taken with the
interference filter but without the etalon, to correct for
pixel-to-pixel gain variations in the CCD detector and for vignetting
within the telescope and spectrograph.  The fundamental design of SALT
causes the effective collecting area of the primary mirror to vary
over a track, resulting in an overall sensitivity variation of about
30\%.  Also, the SALT spherical aberration corrector introduces
differential vignetting across the field of view that varies over the
course of a track with an amplitude of approximately 5 to 10\%.  We
use our R-band photometry of NGC 2280 obtained at the CTIO 0.9m
telescope to correct for these latter two effects.  We compare the
photometry of $\sim 45$ stars present in both the SALT and CTIO images
to determine a smooth two-dimensional quadratic polynomial for each
Fabry-Perot exposure that both corrects the residual flat field
variations and normalizes the images to a common sensitivity.  After
correction, the stellar photometry in the Fabry-Perot images is
accurate to 2 to 4\% and is primarily limited by photon statistical
noise in the narrow-band exposures.

\subsection{Line profile}
The transmission profile of the etalon as a function of wavelength is
well-described by a Voigt function,
\begin{equation}
V(\lambda;\sigma,\gamma) = \int_{-\infty}^{\infty}
G(\lambda';\sigma)L(\lambda-\lambda';\gamma)d\lambda',
\end{equation}
where $G(\lambda;\sigma)$ and $L(\lambda;\gamma)$ are Gaussian and
Lorentzian distributions, respectively.  For the medium resolution
etalon used in these observations, there is significant degeneracy
between the widths of the Gaussian and Lorentzian components, $\sigma$
and $\gamma$.  The instrumental profile is fitted well with the
parameters $\sigma = 1.27\;$\AA\ and $\gamma = 1.02\;$\AA, giving a
FWHM of $4.23\;$\AA, or a velocity equivalent of $196\;$km~s$^{-1}$.

\subsection{Wavelength measurements}
Since light that reaches different points in the image illuminates the
etalon at different angles in the collimated beam, the wavelength of
peak transmission varies over the image.  The wavelength of peak
transmission at a pixel is given by
\begin{equation}
\lambda_p(z,t,R) = \frac{\lambda_0(z,t)}{\sqrt{1+R^2/F^2}},
\label{fpfunction}
\end{equation}
with
\begin{equation}
\lambda_0(z,t) = A + Bz + Et.
\label{lambdazero}
\end{equation}
Here $z$ is a control parameter describing the spacing of the etalon
plates, $t$ is the time of the observation, $R$ is the pixel's
distance from the optical axis of the etalon, $F$ is the focal length
of the camera optics measured in units of image pixels, and
$\lambda_0$ is the wavelength of peak transmission along the optical
axis.  As seen in equation (\ref{lambdazero}), $\lambda_0$ varies
linearly with $z$ (we find no evidence for a more complicated
dependence), and we also allow for a linear drift of $\lambda_0$ with
$t$.  The small fitted value $E\sim2\;$\AA~hr$^{-1}$ is a measure of
the stability of the instrument.

Across a single exposure, $\lambda_p$ depends only on $R$.  A
uniformly illuminated, monochromatic source will therefore be imaged
as a ring centered on the optical axis of the etalon.  This ring's
intensity will be greatest at the radius $R_{\rm line}$ for which
$\lambda_p(z,t,R_{\rm line}) = \lambda_{\rm line}$.  Our 50 exposures
of NGC 2280 contain several rings corresponding to night-sky emission
lines of OH and [\ion{N}{2}] \citep{Oste96}.  We use these rings to
calibrate the four coefficients $A$, $B$, $E$, and $F$.  We fit a
separate wavelength solution for each of the two nights of
observations.  Using $\sim 15$ imagings of 4 night-sky emission lines
from each night, we fit the parameters $A$, $B$, $E$, and $F$ such
that the root mean square residual to our fit is minimized.  The root
mean square residuals for our two wavelength solutions are
$\lambda_{\rm rms}=0.159\;$\AA\ and $\lambda_{\rm
  rms}=0.089\;$\AA\ for the 1 Nov 2011 and 28 Dec 2011 data
respectively.

For the data taken on 1 Nov 2011, we find that the neon calibration
lines systematically deviate from our best fit to the night-sky
emission lines by $\sim 3\;$\AA\ and we therefore use only the
night-sky lines to calibrate our wavelengths for that night.  We do
not understand the large discrepancy between night sky and calibration
system lines, but similar differences have been found in other RSS FP
data sets, and the current SALT calibration system is known to
illuminate the telescope's spherical aberration corrector differently
than does the primary mirror.  The neon calibration lines from 28 Dec
2011 are consistent with the night-sky emission lines and we include
both sets of lines in our wavelength solution for that night.

Once the night-sky emission lines have been used for wavelength
calibration, they must be removed since they often overlap the image
of NGC 2280.  We compute a Voigt function centered around each line
and map these functions onto our images by inverting equation
(\ref{fpfunction}).  We then subtract these rings from our images,
effectively removing the night-sky emission.

\subsection{Ghosts}
Reflections between the CCD detector and the Fabry-Perot etalon give
rise to a ghost image, termed the ``diametric ghost'' in the
discussion of \citet{Jone02}.  This ghost image is in focus and is
mirror-symmetric about the optical axis, with an amplitude of
approximately 5-8\%.  The amplitude of the ghost image depends upon
the quantum efficiency and reflectivity of the CCD, the etalon
transmission and reflection profiles, the filter transmission curve,
and the spectrum of the source; due to these complexities, the ghost
amplitude varies over the image.  We determine the precise position of
the optical axis in our images by measuring the positions of $\sim20$
bright stars and their ghosts.  We rotate each image about this axis
by 180$^\circ$, scale by a factor of 0.065, and subtract this ``ghost
image'' from the original.  While this procedure overcorrects some
stellar ghosts and undercorrects others {by as much as 1.5\%}, it
is effective at removing the reflection of the galaxy's core, where
the ghost would most affect our measurements.  We hope to implement
calibration procedures in the future that will more accurately correct
for the diametric ghost.

\subsection{Velocity measurements and uncertainties}
We use the measured centroids of $\sim 35$ stars present in our SALT
data to align our images such that the same pixel corresponds to the
same point on the sky in all 50 exposures.  The precision of the
individual stellar centroids is $\sim 0.1\arcsec$.  The required
shifts are $\sim 0.1\arcsec$ between images taken during a single
night, but we also had to rotate our images from the second
night by $110.85^\circ$.

In order to improve signal-to-noise (S/N), we employ a $9\times9$
pixel binning in each FP frame.  Some groups \citep{CC03, Epin08} use
Voronoi tesselation to combine low S/N pixels until the spectral line
shift can be measured reliably.  While this procedure indeed allows
velocities to be fitted in regions where the line signal is weak, we
have not adopted it here.  The principal science goal of this study is
to determine the rotation curve; adding more low S/N points to our
maps will have very little effect on the fit, as the extra points will
have low weight in the fit.  Furthermore, the variable spatial
resolution of such a map would greatly complicate the extraction of a
rotation curve, as one should take into account the differing
uncertainties in the locations of the measured velocities.

\begin{figure}
\begin{center}
\includegraphics[width=.9\hsize]{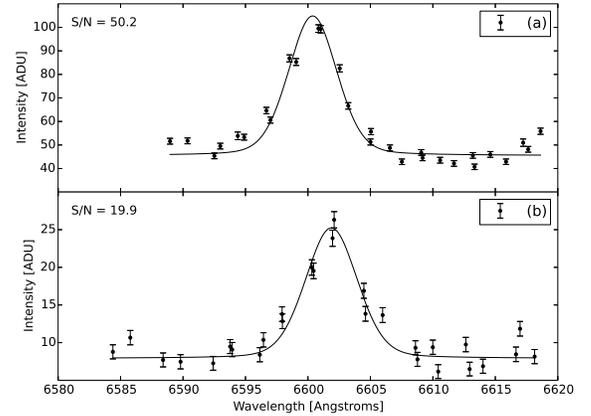}
\end{center}
\caption{Sample spectra and Voigt profile fits to individual pixels in
  our H$\alpha$ velocity map data. The pixels whose profiles are
  plotted here are marked in Figure~\ref{fig:fpdata}.
\label{fig:spectra}}
\end{figure}

\begin{figure*}
\begin{center}
\includegraphics[width=.8\hsize]{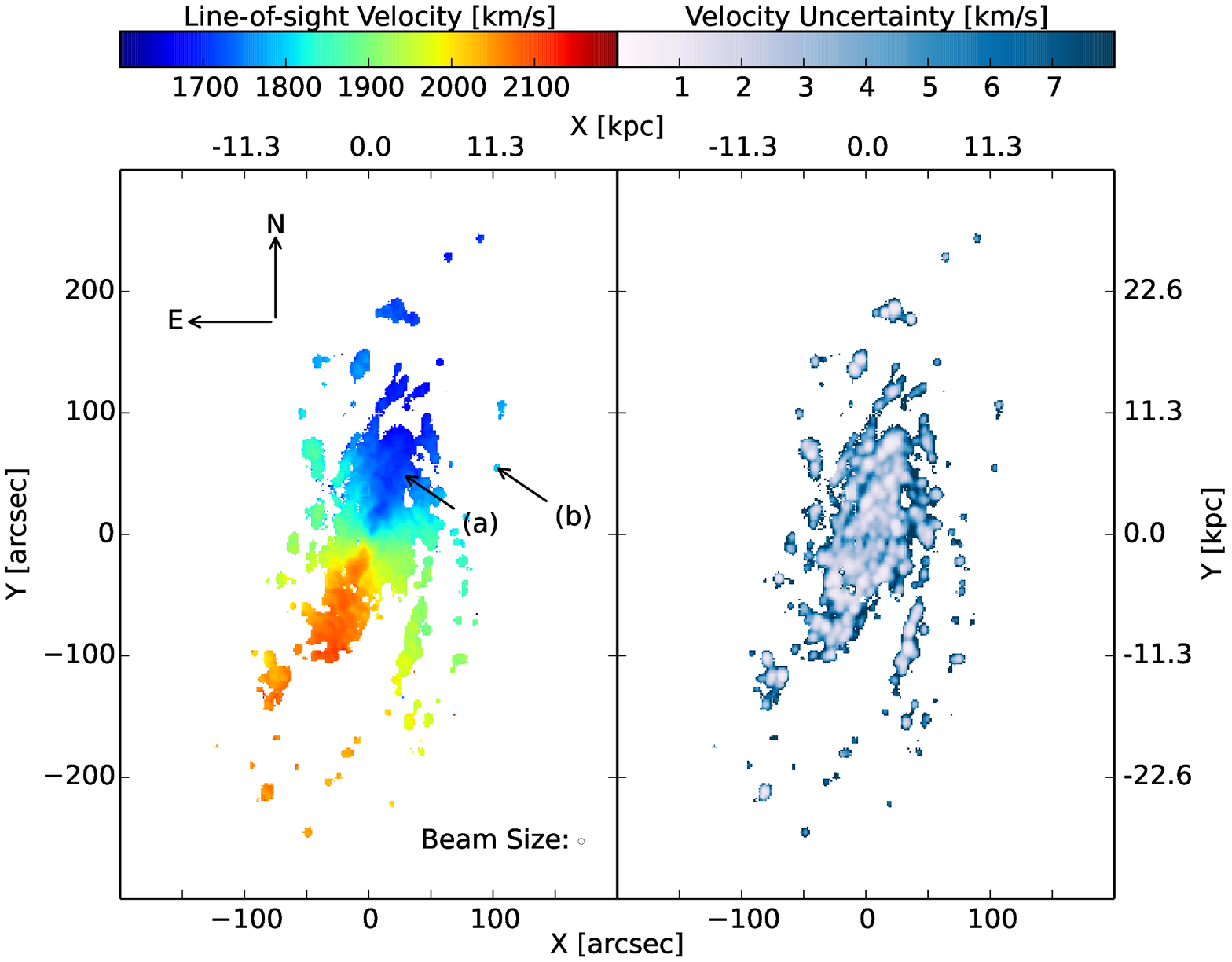}
\end{center}
\caption{Velocity map (left) and velocity uncertainties (right)
  produced from the H$\alpha$ data.  The small circle in lower right
  corner of the velocity map represents the 4.9\arcsec effective beam
  size. Arrows indicate the pixels whose line profiles have been
  plotted in Figure~\ref{fig:spectra}.
\label{fig:fpdata}}
\bigskip
\begin{center}
\includegraphics[width=.8\hsize]{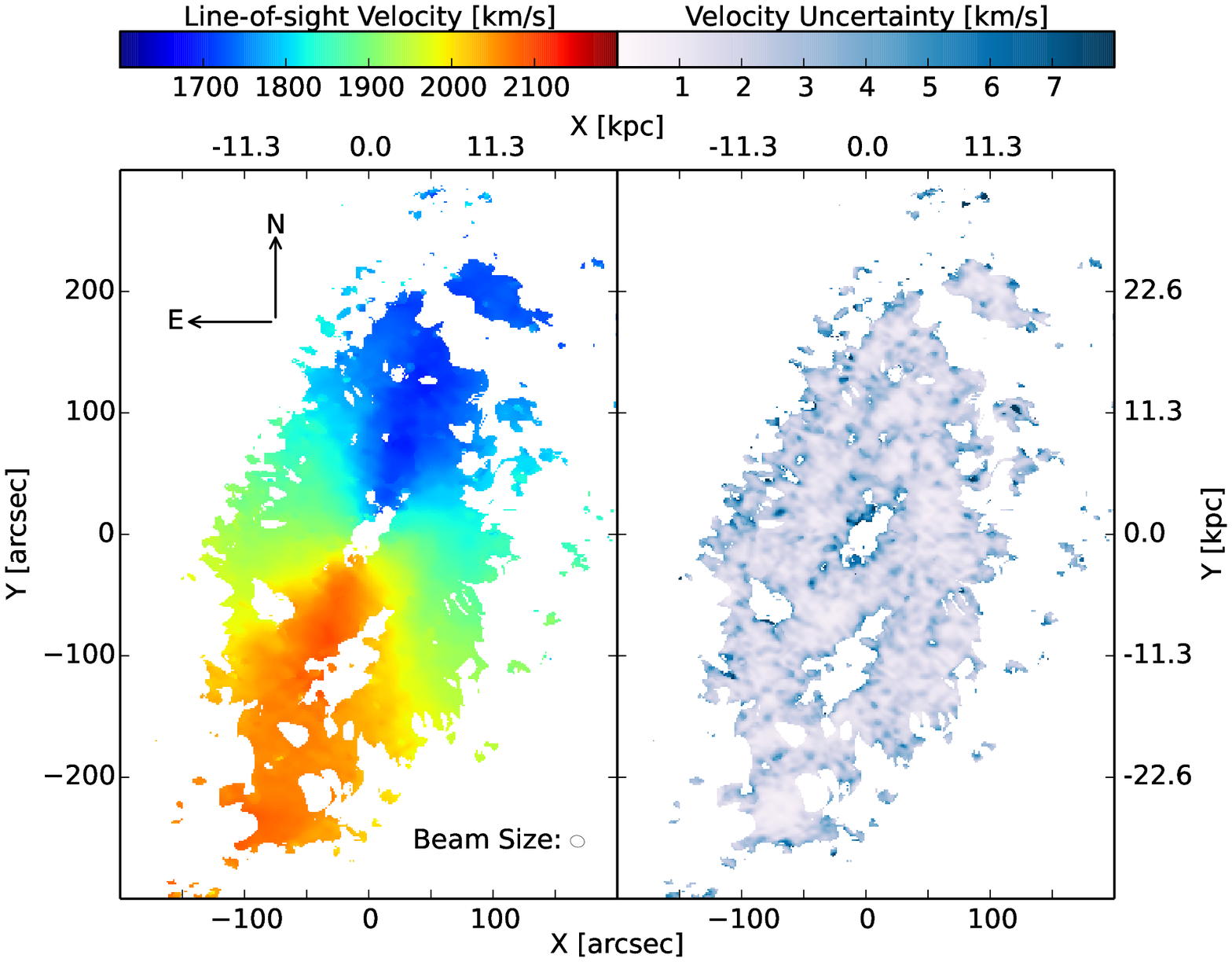}
\end{center}
\caption{Velocity map (Left) and velocity uncertainties (Right)
  produced from the \ion{H}{1} data.
\label{fig:radiodata}}
\end{figure*}

Once aligned, the data sets from the two nights were combined and
fitted simultaneously.  The combined series of images taken at
different values of $z$ provides a short piece of spectrum at every
point.  Etalon spacings were chosen such that the values of
$\lambda_0(z,t)$ were spaced $\sim 2\;$\AA\ apart.  We fit the
spectrum at each pixel with a Voigt function to determine the
wavelength of peak H$\alpha$ emission and the uncertainty in that
wavelength.  In general, we fit for five parameters: the line strength
$I$, the continuum strength $I_0$, the line center $\lambda_{\rm
  peak}$, the Gaussian width $\sigma$, and the Lorentzian width
$\gamma$, together with uncertainty estimates in these values.  We
assume that the H$\alpha$ emission profile is purely Gaussian, and the
Lorentzian width $\gamma$ is held fixed at the measured value from the
neon calibrations.  The wavelength $\lambda_{\rm peak}$ of peak
emission and the uncertainty in determining this peak wavelength
$\Delta \lambda_{\rm peak}$ are then used to calculate a velocity and
velocity uncertainty for each pixel:
\begin{mathletters}
\begin{equation}
\label{eqnv}
v = c(\lambda_{\rm peak}-\lambda_0)/\lambda_0
\end{equation}
\begin{equation}
\label{eqnsigma}
\Delta v = c(\Delta \lambda_{\rm peak})/\lambda_{\rm peak}
\end{equation}
\end{mathletters}
Velocity uncertainties for pixels with strong H$\alpha$ emission are
typically $\sim 3\;$km~s$^{-1}$.  We use the ratio of line strength to
its uncertainty, $I/\Delta I$, as a proxy for S/N and discard velocity
measurements from all pixels with values of $I/\Delta I<5$ from our
velocity map.  We have examined individual spectra of several hundred
pixels and have found that they are each well-fitted by a single Voigt
function.  However, any possible spectral features such as multiple
peaks that are more closely spaced than our spectral resolution
$4\;$\AA\ would be undetectable.  Two sample spectra and Voigt profile
fits for representative high and low S/N pixels are presented in
Figure~\ref{fig:spectra}.

We also detect emission at $6583\;$\AA\ from [\ion{N}{2}] in NGC 2280,
but as our wavelength coverage was not complete enough to capture the
full line at all velocities, we are unable to produce a second
velocity map from the [\ion{N}{2}] data.  However, where the line
strength and wavelength coverage permitted, the fitted velocities
using the [\ion{N}{2}] line are consistent with those from the
H$\alpha$ line.  When fitting the H$\alpha$ line, we exclude
wavelengths more than $13\;$\AA\ redward of its peak so as to fit only
the H$\alpha$ emission and continuum.  \Ignore{Note also that the
  [\ion{N}{2}] 6548 line is marginally visible in the spectrum of
  Figure (2).  This is the weaker of the two [\ion{N}{2}] lines, so
  this strength is consistent.}

Our H$\alpha$ velocity map comprises $\sim4.5\times 10^4$ pixels which
meet our S/N threshold of $I/\Delta I>5$; of these 950 are completely
independent.  Figure~\ref{fig:inty}(c) shows the distribution of
H$\alpha$ emission in NGC 2280.

\begin{deluxetable}{lc}
\tabletypesize{\footnotesize}
\tablewidth{0pt}
\tablecolumns{3}
\tablecaption{\ion{H}{1} Observation Parameters\label{table:HI}}
\tablehead{\colhead{Parameter}&\colhead{Value}}
\startdata
VLA configuration         & BC\\
Synthetic beam FWHM       & 11.46\arcsec $\times$ 9.54\arcsec at 74.39$^{\circ}$\\
Time on-source            & 370 minutes\\
Usable total bandwidth    & 2.5 MHz\\
Band center (helio)       & 1888 km~s$^{-1}$\\
Spatial resolution        & 10.46\arcsec \\
Spatial resolution        & 1.18 kpc \\
RMS noise, \ion{H}{1} line& 0.47 mJy/beam
\enddata
\end{deluxetable}

The H$\alpha$ velocity and uncertainty maps are shown in
Figure~\ref{fig:fpdata}.  The velocity map is typical of a rotating
circular flow pattern seen in projection, and the close spacing of the
velocity contours near the center is indicative of a steep inner rise
to the rotation curve.  There is little evidence for a twist in the
inner flow pattern, suggesting that any possible bar or oval
distortion in the mass distribution is either very weak, or is aligned
with the one of the principal projection axes.

The velocity field is well-sampled in the inner parts, but is
increasingly sparsely sampled as the distance from the center
approaches 200\arcsec, or $\sim 25\;$kpc.  The distribution of
well-defined velocities for \ion{H}{2} regions is not random and
traces the spiral structure somewhat.  Our $9 \times 9$ binning
procedure to improve S/N gives a characteristic minimum size of
4.9\arcsec\ to the velocity ``blobs'', which is particularly apparent
in the right panel of Figure~\ref{fig:fpdata}, where the velocity
uncertainty increases outward from the center of each blob as the S/N
decreases.

\section{\ion{H}{1} Observations}
\label{radio}
Our \ion{H}{1} observations were taken on 13 Oct 2006 using the VLA in
BC configuration during the expansion phase of the array, which at the
time contained five retro-fitted EVLA antennae.  Other parameters of
the observations are listed in Table~\ref{table:HI}.  Standard flux
calibrators, 3C147 and 3C286, were observed at the beginning and end
of the observing run.  A nearby phase calibrator, 0609-157, was
observed for $\sim$4.5 minutes for every $\sim$30 minutes on the
target source for a total of 370 minutes of on-source observing.

Data editing and reduction was completed using the Astronomical Image
Processing System (AIPS) version 31Dec12 \citep{Grei03}.
Due to technical reports of power aliasing affecting the EVLA-EVLA
baselines, all ten potentially affected baselines were flagged and
removed from the data.  The remaining data were then calibrated using
standard AIPS tasks where the flux density for each source was
determined relative to that of the flux calibrators and the phase of
the data was retrieved from the computed phase closures from the phase
calibrator.  Bandpass solutions were also derived using the flux
calibrators to correct for amplitude variation.  The edited cube was
imaged with robust weighting and cleaned, producing a synthesized beam
FWHM of $11.46\arcsec \times 9.54\arcsec$ at a position angle of
$74.39^{\circ}$, a spectral resolution of $10.4\,$km s$^{-1}$, and an
RMS noise in each spectral channel of $\sigma=0.47\,$mJy.

The differing spatial and spectral resolutions of our \ion{H}{1} and
FP data necessitate different approaches to extracting velocity maps.
Velocity spreads across the beam, known as ``beam smearing''
\citep[\eg][]{Bosm81, Swat00}, can give rise to aysmmetric line
profiles that can be corrected for, to some extent, in the \ion{H}{1}
data with its higher spectral resolution.  Beam smearing is less of a
problem in higher spatial resolution optical data, but could not be
corrected for in our data, as the instrumental broadening effectively
hides any intrinsic asymmetry in the line profiles.  We therefore
adopt a different approach to extract the best possible velocity map
from our \ion{H}{1} data.

\begin{figure}
\begin{center}
\includegraphics[width=\hsize]{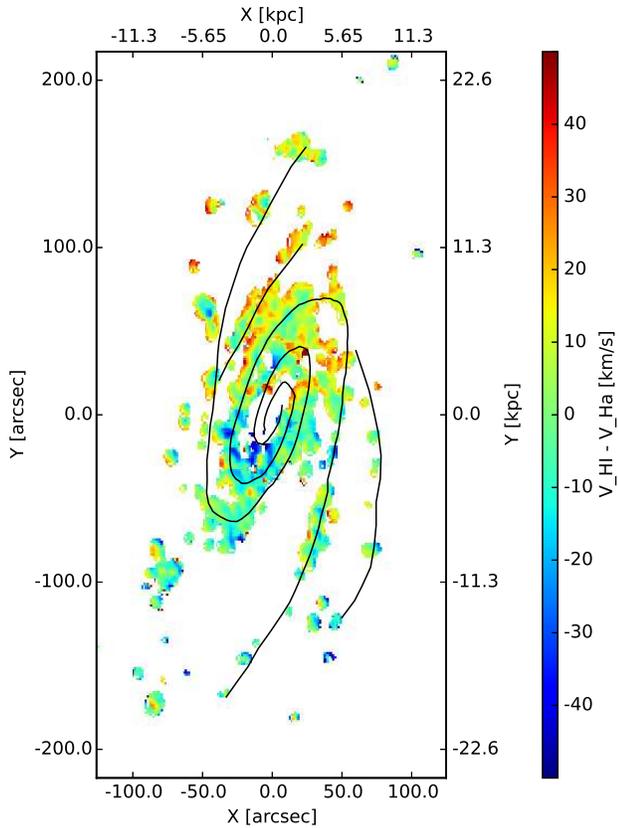}
\end{center}
\caption{Differences in line-of-sight velocity between individual
  pixels in our H$\alpha$ and \ion{H}{1} velocity maps.  The black
  lines show visually-estimated locations of the spiral arm features.
\label{fig:velvel}}
\end{figure}

Figure~\ref{fig:inty}(d) shows a map of the total \ion{H}{1} intensity
$\int I dv$ derived from the continuum-subtracted, calibrated data
cube.  Locations where $\int{I dv}<6.4 \times
10^{20}\,\mathrm{atoms\,cm^{-2}}$ before primary beam correction --
the $3\sigma$, $3$-channel sensitivity limit of the cube -- have been
masked.  We estimate velocities and uncertainties from a Hermite 3 fit
\citep{vdMa93} following a similar approach to that described in
\citet{Noor05} \citep[see also][]{deBl08}.  We use fits obtained from
a smoothed and clipped version of the data cube as initial guesses in
our fit to the full-resolution cube.  The resulting Hermite 3
velocities are kept only for pixels where (a) the uncertainty on the
best fitting velocity is $\sigma_V < 25\,$km s$^{-1}$, (b) the peak of
the Hermite 3 profile exceeds 3$\sigma$, (c) the velocity dispersion
$\sigma_d$ is in the range $5\,$km s$^{-1} < \sigma_d < 100\,$km
s$^{-1}$, and (d) $\int I dv > 0$ in the intensity map of
Figure~\ref{fig:inty}.

Figure~\ref{fig:radiodata} shows the Hermite 3 velocity field and
uncertainties for NGC 2280.  As for Figure~\ref{fig:fpdata}, the
\ion{H}{1} velocity map reveals a similar circular flow pattern, but
with some differences.  Velocities are determined over a wider region,
with more complete coverage in the outer parts, but we were unable to
obtain velocities in an elliptical region at the center that is
extended roughly along the major axis.  Even without data in this
region, the inner velocity gradient appears shallower than in the
H$\alpha$ velocity map.  The flow pattern is remarkably regular, with
no clear twists in the iso-velocity contours, even in the outer parts,
suggesting that any possible warp in the disk of NGC~2280 is mild and
that the galaxy is undisturbed.

\newpage
\section{Velocity Map Comparison}
Figure~\ref{fig:velvel} shows differences between the velocities in
our H$\alpha$ and \ion{H}{1} velocity maps where both measurements are
above our S/N thresholds.  Differences are generally small (green
color), but larger differences of $\ga 40\;$km~s$^{-1}$ are evident in
some places and are not randomly distributed.  In some parts,
particularly in the north, positive differences (red) persist over
regions that are azimuthally extended and there are hints that they
could be associated with spiral features, shown by the black lines,
although many red areas also lie squarely between arms.  There are
fewer differences $\la -40\;$km~s$^{-1}$ (blue) and they show
essentially no correlation with spiral locations.

Note that a very few pixels have absolute velocity differences between
the H$\alpha$ gas and \ion{H}{1} gas ranging up to $\sim
200\;$km~s$^{-1}$.  This is not without precedent, \citet{Zanm08} also
noted a few \ion{H}{2} regions in NGC~1365 having measured velocities
that differed by $60 - 80\;$km~s$^{-1}$ from that of the surounding
gas and \ion{H}{1}, and suggested they could be outside the plane and
seen in projection.

While the loci of positive differences $\sim 40\;$km~s$^{-1}$ do
coincide with spiral arms in one or two instances, interpreting them
as due to unusually strong spiral arm streaming makes little dynamical
sense, for the following reason.

The NW side of the galaxy is approaching, which means that the NE side
is tipped towards us if we assume the spirals trail.  The difference
map shows small velocity differences (green) that suddenly jump to
$\sim 40\;$km~s$^{-1}$ (red) on the NE sides of the spiral arms in one
or two places in the N part of the map.  The sign of the velocity
difference ($v_{\rm H\;I} - v_{\rm H\alpha}$) implies that the
H$\alpha$ emission is coming from gas that is moving, relative to that
producing the \ion{H}{1} line, towards us along our line of sight.
Since the NE is the near side of the galaxy, this excited gas is
therefore moving radially outwards from the center of the galaxy.  (We
discount the alternative possibility that the \ion{H}{1} is moving
away, since there are features in the H$\alpha$ velocity maps at these
locations that are not present in the \ion{H}{1} map.)  However,
spiral streaming theory suggests that post-shock gas should have a
component of velocity in towards the galaxy center inside corotation
for the pattern.  So either the spiral arms are outside corotation
wherever this difference is seen, and the gas acquires its outward
velocity {\em before} it passes through the arm, or the spirals are
leading.  Neither of these alternatives seem palatable.

Since the velocity differences are not well correlated with spiral
arms, nor do they have the sign expected from spiral streaming theory
where they do coincide with spirals, we suspect that they have some
other origin.  We note that similar patchy differences between $v_{\rm
  H\;I}$ and $v_{\rm H\alpha}$ were reported by \citet{Phoo93} in the
galaxy NGC~4254, for which they could find no convincing explanation.

We use the biweight \citep{Beer90} to estimate the Gaussian spread of
the velocity differences.  This estimator is designed to determine the
spread from the bulk of the data without being biased by possible
heavy tails in the distribution.  Using the biweight, we find the mean
velocity difference between our two maps to be $v_{\rm H\;I}-v_{\rm
  H\alpha}=2.5\;$km~s$^{-1}$ and the 1-$\sigma$ scatter in these
diifferences is $14.3\;$km~s$^{-1}$ about the mean.  We find lower
dispersions, $12.0\;$km~s$^{-1}$ for the approaching side only, and
$11.9\;$km~s$^{-1}$ for the receding side, consistent with minor
systematic differences for the two sides discussed below.  These last
two values are entirely consistent with intrinsic ISM turbulent
motions of $\sim 8.0\;$km~s$^{-1}$ \cite{Gunn79}, combined with a
difference of two observed quantities with our (small) estimated
uncertainties.  Thus aside from the patchy large differences shown in
Figure~\ref{fig:velvel}, our two velocity measurements in NGC~2280
generally differ by an amount that is no more than the usual
expectation for turbulence in the ISM.

\section{Velocity Map Fitting}
\label{discussion}
The \textit{DiskFit}\footnote{\textit{DiskFit} is publicly available
  for download at http://www.physics.rutgers.edu/~spekkens/diskfit/}
software package \citep{Spek07, SeZS10} is ideally suited for analysis
of velocity maps such as ours.  In its simplest version, the program
constructs a model of a circular flow pattern in an inclined, thin,
flat disk.  It adjusts the circular speed at a specified set of radii,
together with the systemic velocity, disk center, inclination and
position angle of the entire disk, to find the values of these
parameters that minimize the net $\chi^2$ difference between the model
flow and the velocity map.  It is also capable of fitting non-circular
flow patterns caused by bars, or radial variations to the projection
angles caused by warps.  Unlike other methods, it therefore utilizes
all the data in the entire map to constrain the projection geometry,
and uses a bootstrap technique to yield meaningful uncertainties in
all the fitted parameters.

We have employed this software to fit our H$\alpha$ and \ion{H}{1}
velocity maps of NGC 2280.  For each of our two velocity maps, we fit
two separate models: one for which the position angle, ellipticity,
and center were free to vary, and the other with these parameters held
fixed at common values so that we can directly compare the fitted
rotation curves.  Our first fits are of simple disk models without
bars or warps.

The measurement uncertainties needed to estimate $\chi^2$ for the fit
are the $\Delta v$ found when fitting the H$\alpha$ and \ion{H}{1}
lines.  Since the emission we observe may not be coming from matter
traveling at the circular speed, we add an additional $8\;$km~s$^{-1}$
in quadrature to the uncertainties in both velocity maps.  This value
is typical of turbulent motions in the interstellar medium of a galaxy
\citep{Gunn79} and roughly that estimated above from the scatter in
the differences between our two velocity maps in
Figure~\ref{fig:velvel}.

\begin{deluxetable*}{ccccc}
\tabletypesize{\footnotesize}
\tablewidth{0pt}
\tablecolumns{4}
\tablecaption{ Summary of NGC 2280 Properties \label{table:2280summary}}
\tablehead{\colhead{Property}&\colhead{Previous Work}&\colhead{H$\alpha$ Measurements}&\colhead{\ion{H}{1} Measurements}&\colhead{Favored Values}}
\startdata
\
Position Angle            & $163^{\circ}$\tablenotemark{a} / $155^{\circ}$\tablenotemark{b}     & \fppa  & \radiopa  & \favoredpa  \\
Ellipticity               & $0.56$\tablenotemark{a} / $0.59$\tablenotemark{b}                & \fpeps & \radioeps & \favoredeps \\
Center Right Ascension    & 6h44m49.11s$\pm$0.08s\tablenotemark{c}                           & \fpra & \radiora & \favoredra \\
Center Declination        & -27$^{\circ}$38\arcmin19.0\arcsec$\pm$1.25\arcsec\tablenotemark{c} & \fpdec & \radiodec & \favoreddec \\
Systemic Velocity (helio) & $1895\pm1 \;$km~s$^{-1}$\tablenotemark{d}                         & \fpsys & \radiosys & \favoredsys \\
(cont.)                   & $1899\pm3 \;$km~s$^{-1}$\tablenotemark{e}                         & & & \\
(cont.)                   & $1905\pm6 \;$km~s$^{-1}$\tablenotemark{f}                         & & & \\
(cont.)                   & $2041\pm8 \;$km~s$^{-1}$\tablenotemark{g}                         & & & \\
\enddata 
\tablecomments{A comparison of the previously measured properties of
  NGC 2280 to the properties measured from our best fitting H$\alpha$
  and \ion{H}{1} velocity map models and our favored values.
  Ellipticities from previous works were derived from the major and
  minor axis measurements in those works.}
\tablenotetext{a}{\citet{LV89} from optical photometry}
\tablenotetext{b}{\citet{Jarr03} from 2MASS $K_s$ photometry}
\tablenotetext{c}{\citet{2MASS}}
\tablenotetext{d}{\citet{Spri05} from \ion{H}{1} spectrum}
\tablenotetext{e}{\citet{Kori04} from \ion{H}{1} spectrum}
\tablenotetext{f}{\citet{Bott90} from \ion{H}{1} spectrum}
\tablenotetext{g}{\citet{Sand78} from optical spectrum}
\end{deluxetable*}

\subsection{Projection parameters}
Table~\ref{table:2280summary} gives the fitted values, together with
our estimated uncertainties, of the projection parameters from the two
maps separately.  The two estimates of the systemic velocity differ by
$1.8\;$km~s$^{-1}$, which is well within the quoted 1-$\sigma$
uncertainties.  Our values are smaller than those quoted by
\citet{Kori04} and by \citet{Spri05}.  Because our systemic velocity
was derived from 2-D velocity maps and theirs from a single-dish
\ion{H}{1} measurements, we do not find this difference worrisome.

\begin{figure*}
\begin{center}
\includegraphics[width=.8\hsize]{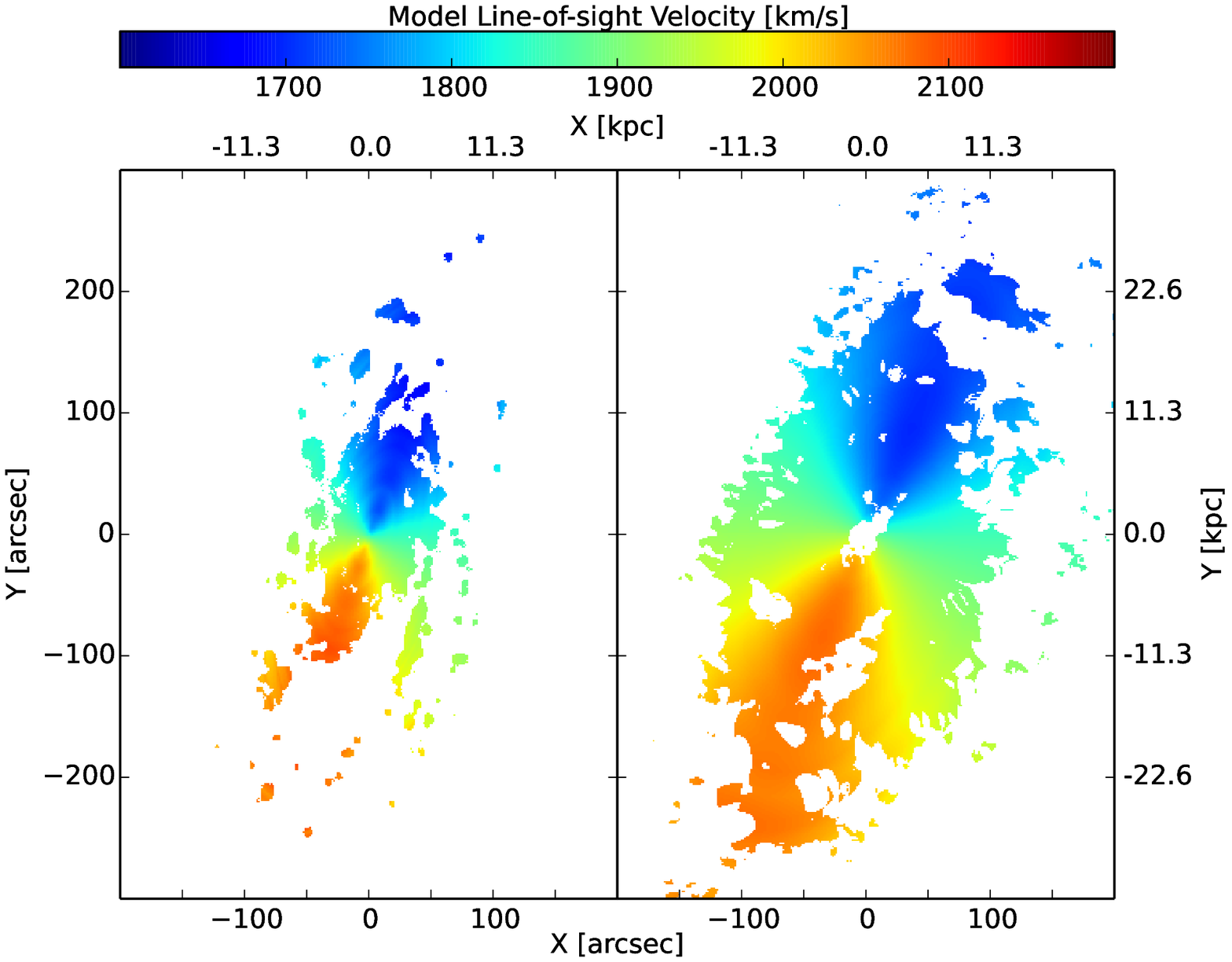}
\end{center}
\caption{Best fitting models to the H$\alpha$ (Left) and \ion{H}{1}
  (Right) velocity maps.
\label{fig:models}}
\bigskip
\begin{center}
\includegraphics[width=.8\hsize]{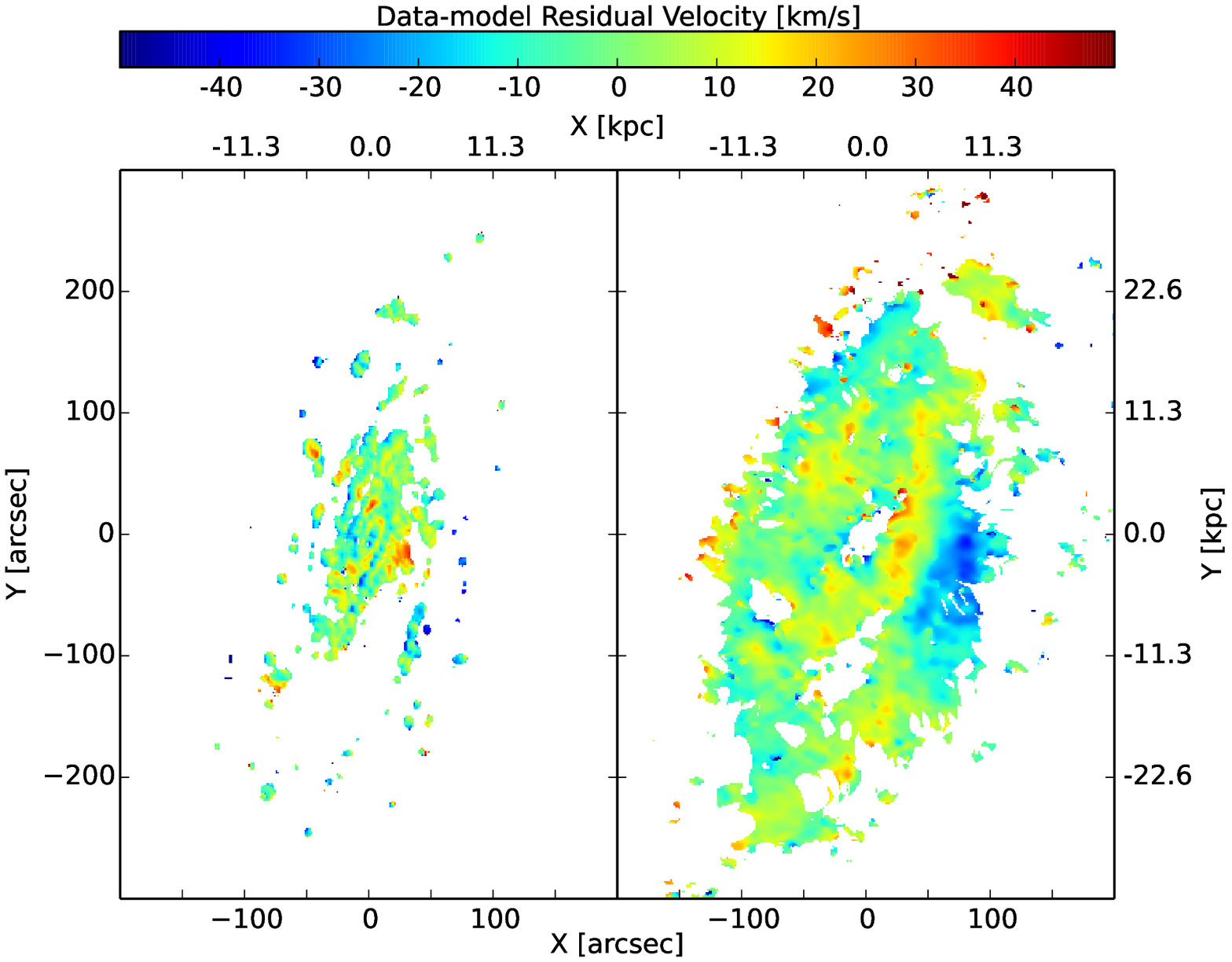}
\end{center}
\caption{Data-model residuals for the best fitting models to the
  H$\alpha$ (Left) and \ion{H}{1} (Right) velocity map.
\label{fig:resid}}
\end{figure*}

Our kinematic estimates of the position angle and ellipticity from our
two maps are also in excellent agreement with each other, and agree
tolerably well with previous apparent ellipticity ($e=1-b/a$)
estimates from a photometric B-band image \citep{LV89} and from 2MASS
near-infrared data \citep{Jarr03} given in
Table~\ref{table:2280summary}.  Mild discrepancies between photometric
and kinematic estimates of inclination can arise from intrinsic mild
ellipticities caused by outer spiral arms, for example.  Our favored
values for the parameters in Table~\ref{table:2280summary} were
determined by weighted means of our two best fitting models, with
weights given by the inverse variances of these quantities.

\subsection{Uncertainty estimates}
Figures~\ref{fig:models} and \ref{fig:resid} show respectively our
best fitting rotation-only models for both the H$\alpha$ and
\ion{H}{1} velocity maps and the corresponding residual maps.  The
values of the reduced $\chi^2$ for these models are 2.33 and 1.75
respectively, indicating a poor formal fit to the data.  The residual
velocities are significantly larger than our typical velocity
uncertainties, which are usually dominated by our adopted value of
$8\;$km~s$^{-1}$ for turbulent motions in the ISM, and are generally
spatially correlated, indicating that the gas in NGC 2280 has more
complicated motion than is allowed for in our models.  For example,
our models do not include non-circular streaming motions associated
with spiral arms.  A dynamical origin for the larger residuals is
supported by the fact that their general pattern is similar in the two
separate datasets.

The {\it DiskFit} program is capabale of fitting a bar flow pattern of
any amplitude and/or a crude model for a warp.  We have attempted to
fit our velocity maps with models that include bars or warps, but
find that the inclusion of these features does not substantially
improve the fits.  However, {\it DiskFit} does not have the capability
to fit more complicated spiral streaming patterns, which are generally
less clearly rotationally symmetric than are bars and probably arise
from superposed multiple spiral modes \citep{SC14}.  Since we are
unable to model these features in our data, we must assess the extent
to which their presence affects the values of our derived quantities.

The usual bootstrap technique adds randomly rearranged residuals at
every pixel to the best-fit model to create many pseudo-data sets, and
determines uncertainties in the parameters from their scatter in fits
to the pseudo-maps.  This approach assumes that the residuals from the
original fit are uncorrelated and, were we to apply it in our case, it
would destroy the correlated nature of the residuals and lead to
uncertainty estimates that would be far too small.  \textit{DiskFit}
therefore offers two distinct methods to retain the correlated nature
of the residuals: since we see that the residuals reflect a spiral
flow pattern, we utilize the method developed by \citet{SeZS10}.  Our
uncertainties are estimated from 1000 bootstrap iterations.

\begin{figure}
\begin{center}
\includegraphics[width=\hsize]{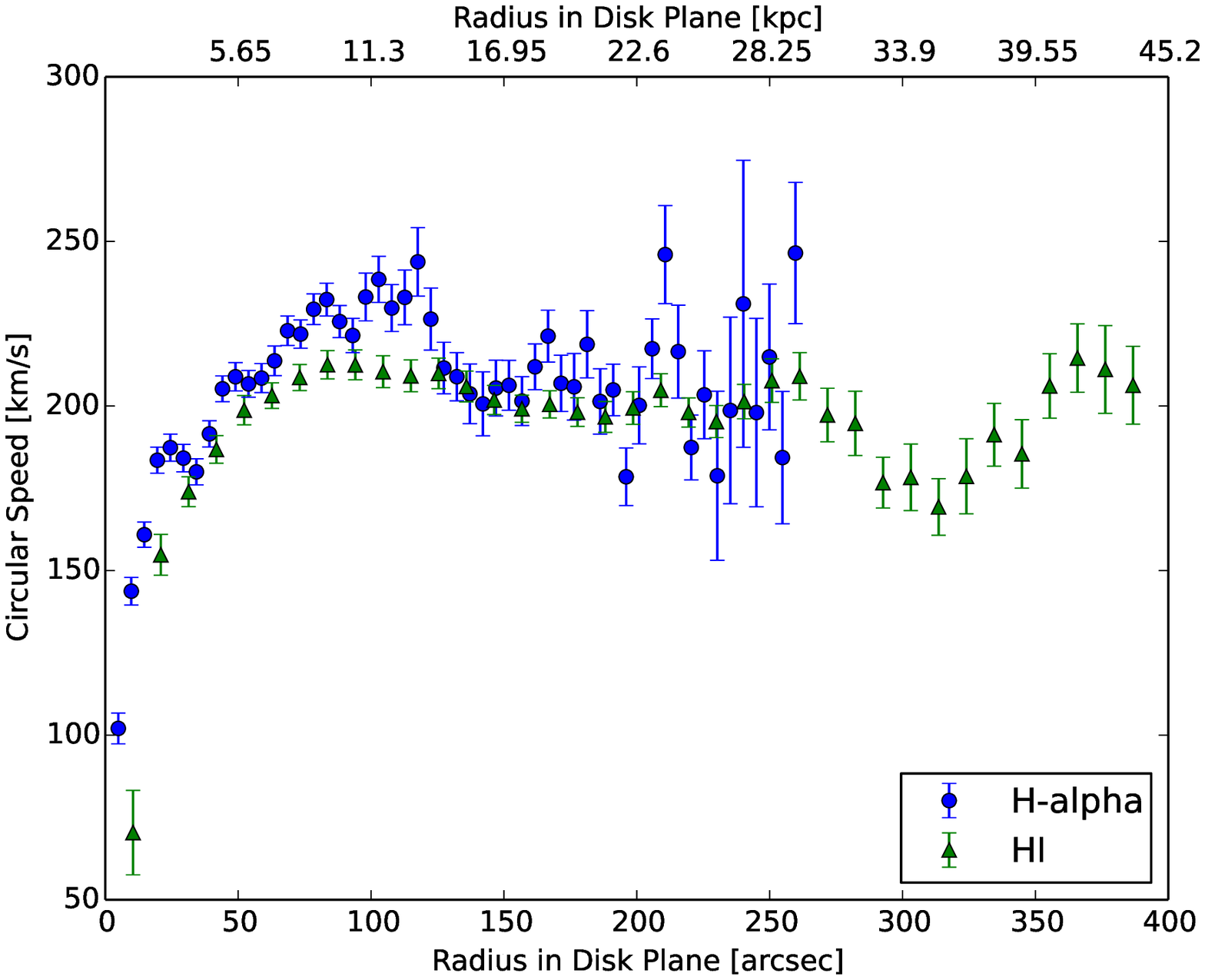}
\end{center}
\caption{Rotation curves for NGC 2280 derived from the model fits to
  our H$\alpha$ and \ion{H}{1} data.
\label{fig:rotcurves}}
\end{figure}

\subsection{Rotation curve}
Figure~\ref{fig:rotcurves} shows the circular speed derived from our
fits to the H$\alpha$ and \ion{H}{1} velocity maps, with the error
bars showing $\pm1\sigma$ uncertainties.  The abscissae were chosen
such that the points are spaced 1 beam FWHM apart.

\begin{figure}
\begin{center}
\includegraphics[width=\hsize]{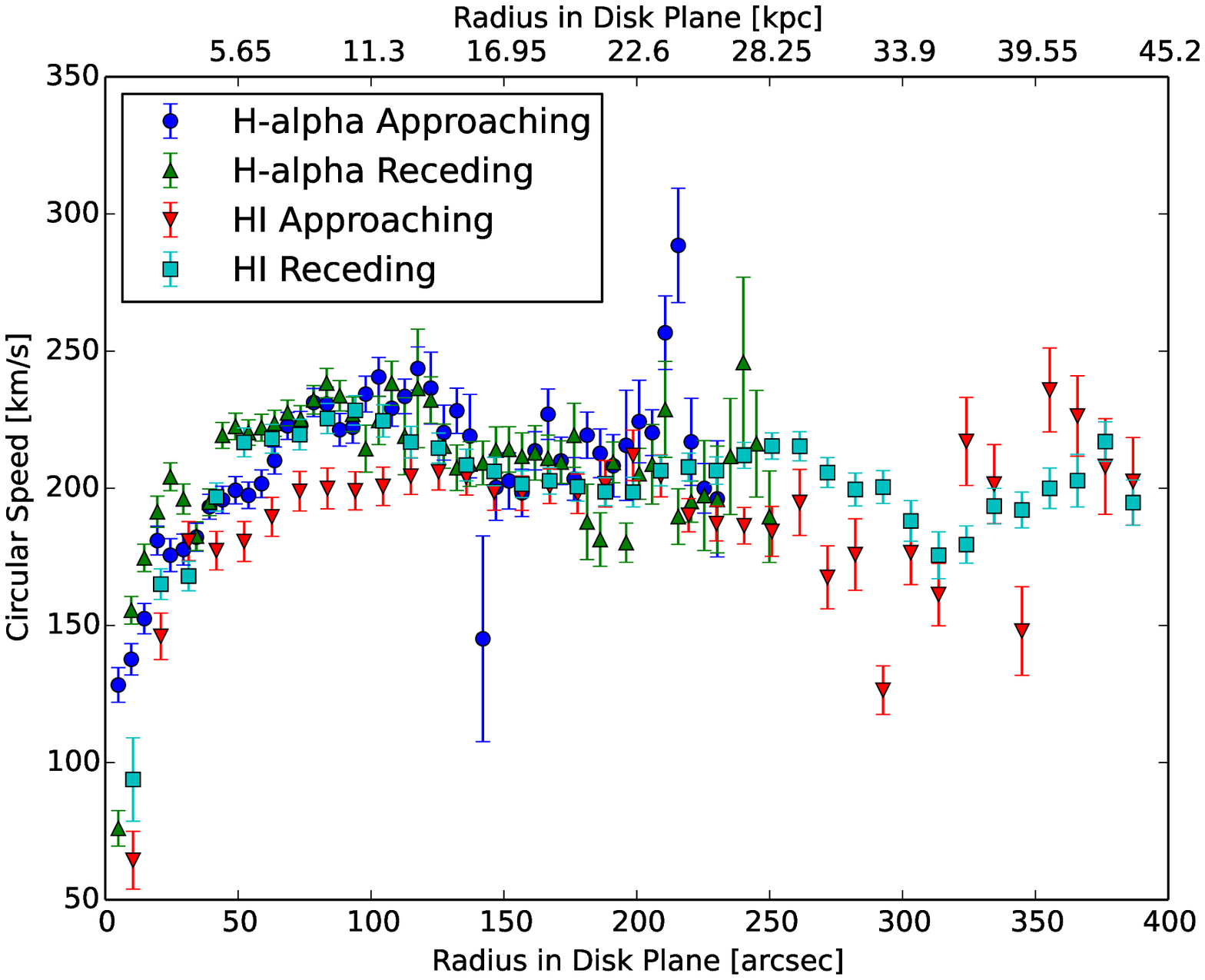}
\end{center}
\caption{Rotation curves for NGC 2280 produced from independent model
  fits to the approachingand receding halves of our H$\alpha$ and
  \ion{H}{1} data.
\label{fig:splitcurves}}
\end{figure}

We have also fitted models to the approaching and receding halves of
the galaxy independently.  For these models, we fit only points which
are north or south of the minor axis, respectively, and we fix the
systemic velocity, position angle, and ellipticity to the favored
values listed in Table~\ref{table:2280summary}.  In
Figure~\ref{fig:splitcurves}, we show rotation curves derived from
these models to both the H$\alpha$ and \ion{H}{1} velocity data.  We
note that the rotation curves produced from these models are largely
consistent with each other.  The extreme outlying points in the
approaching H$\alpha$ curve are determined by very few pixels that are
far from the major axis, and thus are sensitive to deprojection
errors.  There are some asymmetries in the estimated speeds that are
most pronounced over the range $30\arcsec < r < 100\arcsec$.  For both
the H$\alpha$ and \ion{H}{1} velocity maps, the receding side of the
galaxy appears to have a slightly higher rotational speed at these
radii than does the approaching side.  Asymmetries in these fitted
speeds are probably caused by $m=1$ distortions in the galaxy,
although $m=3$ could not be ruled out.  Their presence is somewhat
surprising in a galaxy that seems so undisturbed and hosts such a
regular bi-symmetric spiral pattern.

The combined H$\alpha$ rotation curve (Figure~\ref{fig:rotcurves})
manifests a series of small ``bumps'' that are less prominent or
non-existent in the \ion{H}{1} rotation curve.  These bumps stand out
more clearly on the approaching side (blue circles in
Figure~\ref{fig:splitcurves}), while the rotation curve is smoother on
the receding side.  They are almost certainly related to the large
velocity differences visible in Figure~\ref{fig:velvel}, which
revealed patches of excited gas moving more rapidly towards us (red)
on the approaching side (north west), while there are fewer anomalous
velocties on receding side.  The origin of these kinematic differences
between the excited and neutral gas in NGC~2280 is unclear at this
point.

Aside from these anomalous H$\alpha$ velocities, the rotation curves
derived from our two maps appear to be in reasonable agreement with
each other, and most differences stem from the different spatial
coverage and resolutions of the separate maps, which complement each
other well.  As shown in Figures~\ref{fig:fpdata} and
\ref{fig:radiodata}, the H$\alpha$ velocity map is more sparsely
sampled at large radii than is the \ion{H}{1} velocity map, leading to
larger fluctuations and greater residuals in the optically estimated
circular speed (Figure~\ref{fig:rotcurves}).  However, the inner rise
of the rotation curve is better resolved in the optical data, as we
discuss in more detail next.

\subsection{Inner velocity gradient}
The slope of the inner rise of the \ion{H}{1} circular speed is
shallower than that found from the optical data.  One well-known
reason for the shallower central gradient from the \ion{H}{1}
measurements is beam smearing \citep{Bosm81, Swat00}, which has the
effect of combining velocities of gas away from the major axis with
those from gas on the major axis.  Since projection causes the
line-of-sight velocity to be systematically smaller away from the
major axis, beam smearing always biases rotational velocity
measurements towards lower values.

\begin{figure}
\begin{center}
\includegraphics[width=\hsize]{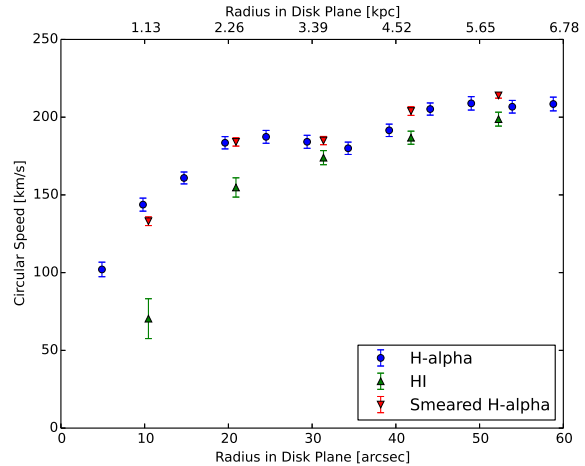}
\end{center}
\caption{Inner rise of rotation curves derived from the model fits to
  our unsmeared H$\alpha$, \ion{H}{1}, and H$\alpha$ smeared to the
  resolution of the \ion{H}{1} data.
\label{fig:zoomedcurves}}
\end{figure}

\begin{figure}
\begin{center}
\includegraphics[width=\hsize]{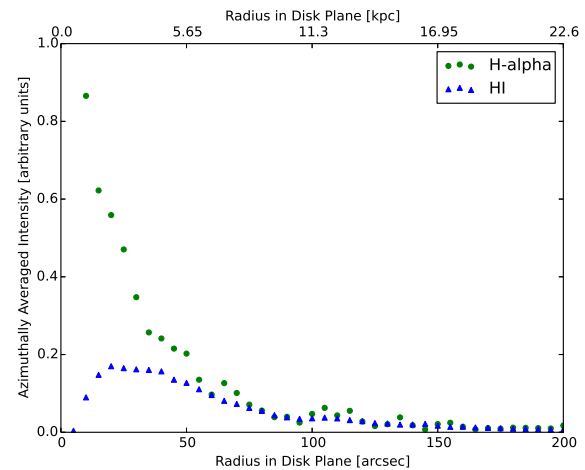}
\end{center}
\caption{Azimuthally averaged intensity profiles of NGC 2280 from our
  H$\alpha$ and \ion{H}{1} data, in $5\arcsec$ radius bins.  Note the
  central hole in the \ion{H}{1} intensty profile that is not present
  in the H$\alpha$ profile.
\label{fig:intyplot}}
\end{figure}

In order to determine whether spatial resolution alone is responsible
for the different central gradients, we have convolved our FP data
with a Gaussian beam identical to that used for the \ion{H}{1} data.
This procedure required us first to take out the center-to-edge
velocity gradient in the FP frames in order to make a new data cube
with single velocity ``channels'' before blurring.  We then modelled
the resulting velocity map with the \textit{DiskFit} software.  In
Figure~\ref{fig:zoomedcurves}, we plot the inner rise of rotation
curves for our unsmeared H$\alpha$ (blue circles), smeared H$\alpha$
(red inverted triangles), and \ion{H}{1} data (green triangles).  As
expected, the beam smeared H$\alpha$ rotation curve has a shallower
rise inside $R=20$\arcsec\ than that of the unsmeared H$\alpha$
rotation curve.

However, the beam smeared H$\alpha$ rotation curve still rises more
steeply than does the \ion{H}{1} rotation curve, although the
\ion{H}{1} data are sparse within $\sim 30\arcsec$ of the galaxy
center.  We attribute the yet shallower rise to a deficiency of
\ion{H}{1} emission in the center of NGC 2280.
Figure~\ref{fig:intyplot} shows the azimuthally averaged intensity
profiles for the H$\alpha$ and \ion{H}{1} data.  The \ion{H}{1}
surface density drops toward the center, consistent with the
$\sim30\arcsec$ central hole in the intensity map
(Figure~\ref{fig:inty}), a common feature of \ion{H}{1} emission near
the center of galaxies \citep[\eg][]{Wang14}.

The weak \ion{H}{1} line cannot yield reliable velocities, whereas the
H$\alpha$ intensity rises continuously into the center enabling
reliable velocity measurements at small radii.  This difference in the
radial distribution is probably responsible for the remaining
difference between the rotation curves.  Note that there must be some
hydrogen in the center to produce the H$\alpha$ emission, and the
neutral gas is probably molecular therefore; however, we have been
unable to find any existing CO observations to confirm the presence of
molecular gas.

\section{Summary}
\label{summary}
We have presented detailed velocity maps of the nearby spiral galaxy
NGC~2280 produced from Fabry-P\'erot measurements of the H$\alpha$
line of excited gas and from aperture synthesis measurements of the
21cm line of \ion{H}{1}.  Despite the fact that these two maps were
derived from different components of the ISM having different spatial
distributions, and were measured with different instruments having
widely differing spatial and spectral resolutions, the measured
velocities are generally in excellent agreement.

The kinematic maps of NGC~2280 reveal a remarkably regular flow
pattern, with no evidence for a bar or oval distortion, or for a warp.
In particular, the galaxy appears to be undisturbed, despite the
presence of numerous, possibly-close companions listed in
Table~\ref{table:neargals}.  We have obtained estimates of the
systemic velocity, disk inclination and position angle using {\it
  DiskFit}.  Our derived values from the two maps agree within their
estimated uncertainties and we favor $\phi_{\rm PA}=$\favoredpa,
$e=$\favoredeps, and $v_{\rm sys}=$\favoredsys.  These values are
broadly consistent with previous estimates, as shown in
Table~\ref{table:2280summary}.

Additionally, the two datasets yield similar rotation curves.  The
differences in spatial extent and resolution between the H$\alpha$ and
\ion{H}{1} velocity maps demonstrate the complementarity of the
measurements.  At small radii, the resolution of the H$\alpha$
velocity map allows for a more precise measurement of the inner slope
of the rotation curve.  The better-filled velocity map from the
\ion{H}{1} allows us to measure the rotation curve to large radii with
greater precision than could be obtained from the H$\alpha$ velocity
map.

We have compared the velocities from the two datasets at overlapping
pixels, and find a near Gaussian spread of differences with a
dispersion of $\sim 14\;$km~s$^{-1}$, consistent with an intrinsic
velocity dipersion of $\sim 8\;$km~s$^{-1}$ broadened by the combined
uncertainties from measurements by the two instruments.  However, a
small fraction of the pixels manifest much larger differences, $\ga
\pm40\;$km~s$^{-1}$, whose origin is unclear. 

Future papers in this series will present both optical and 21~cm
  velocity maps and deep optical photometry of all 19 galaxies in our
  RINGS survey.  Our ultimate goal is to fit these data with mass
  models in order to constrain the radial distibution of dark matter
  in each galaxy and to compare that with predictions from
  $\Lambda$CDM cosmology.

\acknowledgements We thank Tad Pryor for many helpful conversations
about the intricacies of Fabry-P\'erot data reduction and the referee
for a detailed report.  This work was supported by NSF grant
AST-1211793 to JAS and TBW.  TBW acknowledges support by the National
Research Foundation of South Africa under grant CSUR87704.  KS
acknowledges support from the Natural Sciences and Engineering
Research Council of Canada (NSERC).  This research has made use of the
NASA/IPAC Extragalactic Database (NED) which is operated by the Jet
Propulsion Laboratory, California Institute of Technology, under
contract with the National Aeronautics and Space Administration.



\end{document}